# Title: Concerns regarding the deterioration of objectivity in molecular biology

**Tomokazu Konishi**
**Akita Prefectural University**
konishi@akita-pu.ac.jp

**Scientific objectivity was not a problem in the early days of molecular biology. However, relativism seems to have invaded some areas of the field, damaging the objectivity of its analyses. This review reports on the status of this issue by investigating a number of cases.**

## History of molecular biology

The field was fostered by researchers in the areas of genetics, physics, and chemistry [1]. In 1944, Schrödinger wrote one of the most influential books of his time and inspired researchers to study life [2]. Their mission can be summarized by Crick's statement: "the ultimate aim of the modern movement in biology is to explain it all in terms of physics and chemistry" [3]. These efforts revealed the functions and structure of the genome. Physicists and chemists should be familiar with the philosophy of science.

Shortly after, biologists joined the field of molecular biology. As DNA sequencing became popular, the amount of data handling increased dramatically, thus changing the scope of studies in both qualitative and quantitative aspects. The analysis of many short fragments of sequences easily exposed the limitations of biologists. Naturally, the participation of informaticians became required; this can be seen as the beginning of bioinformatics, which provides useful tools that use sophisticated mathematical models and algorithms.

However, the communication between informaticians and molecular biologists is often quite difficult. Researchers belong to different societies and have different senses of purpose. While I belong to several societies related to both activities, I rarely meet other researchers attending both types of conferences. This problem may be based on the differences in their backgrounds. Indeed, this would be one of the incommensurable cases of communication beyond different *intelligent frameworks*, introduced by Popper [4].

## Scientific objectivity

This difficulty may encourage the infiltration of relativism into molecular biology. In its original philosophical meaning [5], relativism cannot be avoided in some advanced areas of science; for example, an observation may be explained differently through the special theory of relativity and quantum mechanics. Such differences may cause problems in understanding the universe. However, in the future, we can expect that the theories will be combined; from this point of view, the present state of molecular biology is much more undeveloped, as is verified later.

Without objectivity, the checking process carried out by peers cannot come into effect; this is one of the essential bases of the importance of scientific objectivity [6]. However, opposing opinions in scientific philosophy may exist; for example, Feyerabend insisted on being free from any fixed idea and explained this attitude as "anything goes" [7], which would be one of the most extreme standpoints of relativism [5]. However, science is based on the integration of the contributions of many researchers; if evidence depends on a particular idea or person and cannot be reproduced by others, the knowledge obtained based on such evidence cannot be integrated. Falsifiability [8] is critically important in this context. In fact, very few contemporary scientific philosophers endorse Feyerabend's anarchistic principle (Prof. Elliott Sober, personal communication, 2015).

Mathematical tools based on different ideas will lead to different answers starting from the same set of observations. Another tool would be introduced by using another idea, thus increasing the number of arbitrary options. This is the current situation in the field of molecular biology in which conclusions are dependent on the idea applied for its derivation. Thus, it is clear that such relativism increases the uncertainty of conclusions and hampers the integration of knowledge.

## Case study: microarray technology

Let us examine one such field, the analysis of transcriptomics data. In the late 1990s, microarray technologies enabled the measurement of the expression levels of some 10,000 genes. This was a *terra incognita* of science. Before the technology was available, only a small number of genes could be covered at once. First, we faced a critical problem: how can the levels of expression be compared between measurements? The signals are not recorded in absolute units, and they tend to depend on each measurement. For example, when comparing the raw data of two measurements, there would be a tendency toward a greater intensity at higher levels, and vice versa (Fig. 1). A process that enables comparisons, *normalization*, was required for analyses. Some informaticians provided a group of methods that use a smoothing algorithm called locally weighted scatterplot smoothing (LOESS) [9,10]. The algorithm finds a function for the tendency; the raw data can be leveled using the inverse function. However, it involves a limitation in handling multiple measurements: as the function is valid for the specific combination of two measurements, combining three or more measurements becomes difficult. Such a problem does not occur when the set of measurements share the same data distribution. To achieve this condition, the robust multi-array average (RMA) was developed (Fig. 2) [11], which determines a standard distribution and then replaces all the data with the standard. Although the RMA has been widely used, proper science, such as chemistry or physics, will never allow these two methods, for the following reasons.

Both methods conflict with the requirements of science on two levels. One is the manipulation of data, defined as *falsification*: "manipulating research materials, equipment, or processes, or changing or omitting data or results such that the research is not accurately represented in the research record" [12]. As the term falsification would usually be used to criticize attempts that lead to certain conclusions, this definition may be overly strict; the methods only lead to certain characteristics of data distribution. However, they conflict conclusively at another level in that they damage the falsifiability of the analysis. Those methods intentionally change the data to fulfill their assumptions; therefore, it is impossible to verify whether the assumptions and compensations were appropriate or not. Thus, the objectivity of the knowledge obtained is totally lost.

A falsifiable normalization method was required to establish transcriptome studies within proper science. In general, data analysis requires an appropriate mathematical model, and such a model is constructed on several assumptions. How can the falsifiability of such assumptions be maintained and verified?

A preferred approach would be exploratory data analysis [13], which aims to find a suitable model for the subjected data using the least number of parameters. As each parameter may stand for an assumption, this parsimonious approach helps reduce the number of assumptions. The appropriateness of the model is verified objectively. The final purpose of the analysis lies in finding the hidden structure of the data, that is, how and why the data have been produced as they are.

Fortunately, a suitable model was found for microarray data [14], which have a lognormal distribution that contains a single arbitrary parameter that represents the background of hybridization (Fig. 3). The distribution was observed in any chipset and data; using this model, the raw data could be transformed into quasi-absolute values of z-scores [15]. The appropriateness of the model is verified in each of the datasets. It is not possible to transfer the full signal range; rather, weak signals contaminated by noise are found and removed (Fig. 3B, red).

The hidden structure of transcriptome data was also disclosed. By integrating the biochemical knowledge on the production and degradation of transcripts in a cell, collaborating functions of cellular regulators and genome DNA were modeled on thermodynamics, which has been used to explain many biochemical events (Fig. 4) [16]. Based on this model, the grammar of the genome code, that is, how the transcriptome is formed by the genome, is explained. The lognormal distribution of transcripts was derived from the model. Several other characteristics of microarray data were predicted by the model and were verified using sets of microarray data. Although a decade has passed since this model was proposed, no conflicting results have been reported. Similar to what is observed for any other scientific theories, this is no more than a tentative model. However, it may give better solutions than other easy and convenient views of the transcriptome.

## Case study: RNA-seq

Shortly after the development of next-generation sequencers, a derivative application to count transcripts was produced, RNA-seq, which estimates the expression levels from the counts and compares the results between samples.

RNA-seq was welcomed with huge expectations, such as high accuracy, because it is based directly on the counting of the number of transcripts; moreover, it was expected to be free from complex mathematical models [17]. In a practical sense, however, counting up any item that is distributed lognormally is not an efficient approach because it requires an enormous amount of counting to ensure the sensitivity necessary for transcriptome analysis. Moreover, early methods of normalization used total counts: each transcript count was normalized by dividing it by the total [18]. This also represents a problem, as a sum total is not robust in lognormally distributed numbers; if the top gene changes its level, the total will be significantly altered. Furthermore, some informaticians started to apply RMA, which has been shown to be inappropriate in microarray research, to RNA-seq [18,19]. Although I am not familiar with the terminology, an economist termed such an ill-conceived idea and its acceptance as *zombie* [20]

There was no difficulty in applying the parametric method to RNA-seq data because the model was found from microarray data; it was tested using the many public datasets available at that time [21]. The distribution was actually found from every dataset; however, all the datasets showed an extremely high noise level that reached z-scores of 1. This simply means that 84% of the genes were buried in noise; in contrast, most of the microarray datasets covered z-scores of –1 or –2. The high level of noise is not caused by a shortage of total reads, but may derive from defects in the experimental procedures. Such noisy data cannot escape the *multiplicity of tests* problem (discussed below) unless the range of data affected by it is completely removed. Unfortunately, the expectations of RNA-seq have not been fulfilled. The disastrous state of this expensive methodology likely arises from the zombie analyses, which have helped hide the defects during its development.

Case study: statistical tests

The multiplicity of tests available for statistical analyses may be another issue that informaticians introduced into this field as an inappropriate idea. Transcriptome data may contain thousands of genes, and we test each of the genes for whether they exhibit significantly changed expression levels. Some informaticians started to claim that they cause multiplicity of tests, a family-wise error rate, and that, therefore, the test results have to be compensated [22]. This sounded plausible, and many methodologies were proposed to circumvent the problem [22-25]. Each of them is based on a certain mathematical model that is founded on assumptions. Although the result of a test would be considerably altered by the compensations, the appropriateness of the assumptions was never verified.

In fact, the multiplicity problem only becomes real when false positives occur at a rate that is close to the threshold. According to Bayes' theorem, this is limited when the likelihood of the null hypothesis is quite high. Such a condition would be true for the maintenance of industrial machine tools, for example. However, it would rarely occur in biological experiments; rather, genes change their expression levels seamlessly, and microarrays can accurately detect responses to the tested stresses. Therefore, the likelihood would be quite low; in fact, the distribution of *p*-values estimated in a microarray experiment is deeply skewed, while that of high-likelihood tests is uniformly distributed [26]. Any compensations for *p*-values would simply introduce errors to the results, and should therefore be avoided. It is interesting to note that statisticians can be classified into three schools of thought based on how they address this issue [27], which may arise from the types of problems they face on a daily basis. The native interpretation of a *p*-value is closest to the attitude of Fisher, who established methods of statistical testing while working in an experimental institute, studying crop yields.

## Case study: phylogenetics

Let us examine the possible origin of the setting in which biologists never criticize the inappropriate recommendations of some informaticians. This disconnection seems to occur in a field of phylogeny that estimates the relationships of samples based on sequence data. These data have been analyzed by generating tree-shaped cluster structures. From the mathematical viewpoint, a method of clustering only suggests a possible shape. This is the limitation of hierarchical clustering; it does not possess the objectivity that is required in proper science. However, biologists frequently use the estimated structure to support their conclusions. Such clustering has been performed on estimated distances between sequences [28]. Many methods can be used to estimate the distances and the shape of clusters. Although the methods practically determine the results, they are accepted in parallel as arbitrary options that may have good and bad points. They assume various conditions that have never been verified. Recently, methods that do not use the distances were also introduced [29]; however, their relationships with falsifiability remain the same.

Although each of these methods is built on several ideas regarding evolution, these ideas are no excuse to negate falsifiability. Evolution is certainly difficult to study in proper science, as critical evidence is difficult to obtain, and it cannot be reproduced experimentally. The tools seem to be the products of intelligent games that are enabled by this limitation. In contrast, an objective approach would not depend on a phylogenic tree, as any tree-shaped model is inappropriate for relationships of biological samples in which the genomes have evolved by both vertical and horizontal transfers [30].

In this review, we discussed the deterioration of objectivity in molecular biology. For every analytical method and field of research, a certain methodology will be required to analyze the data, but respect for the purpose of science is required in the preparation of these methodologies. Moreover, critical assessment regarding the given mathematical model is essential to judge if it has problems, such as poor falsifiability.

**Figure legends**

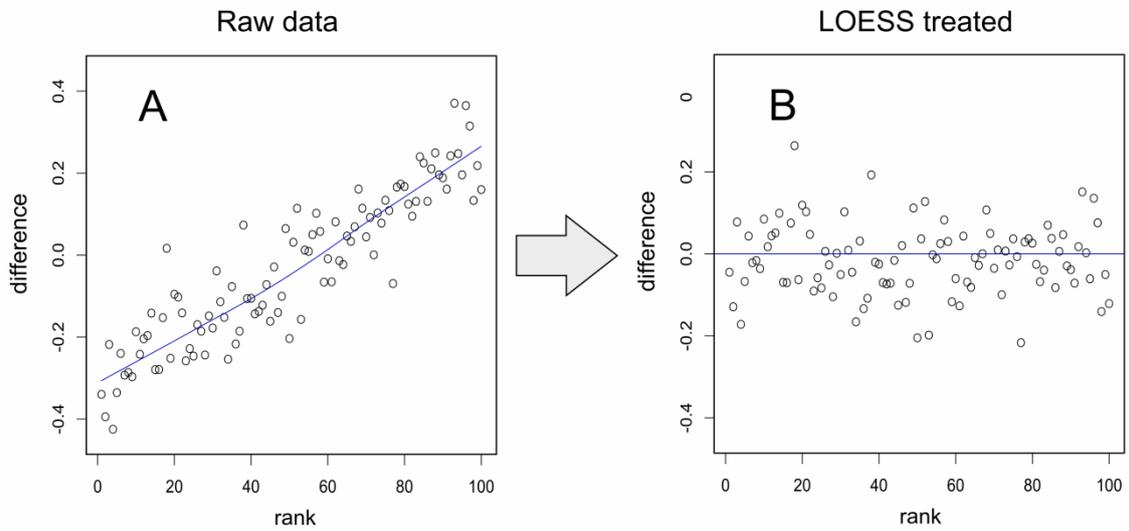

**Fig. 1 Leveling of data using a smoothing function (model)**

**A.** An example of plots comparing two measurements. The differences are presented on the y-axis and the ranks of average levels are presented on the x-axis. A tendency to the ranks may be evident; in this case, differences would become positive for genes that are expressed at higher levels. The tendency could be estimated using a smoothing function, such as LOESS (blue line). **B.** Using the reverse function, the tendency is canceled.

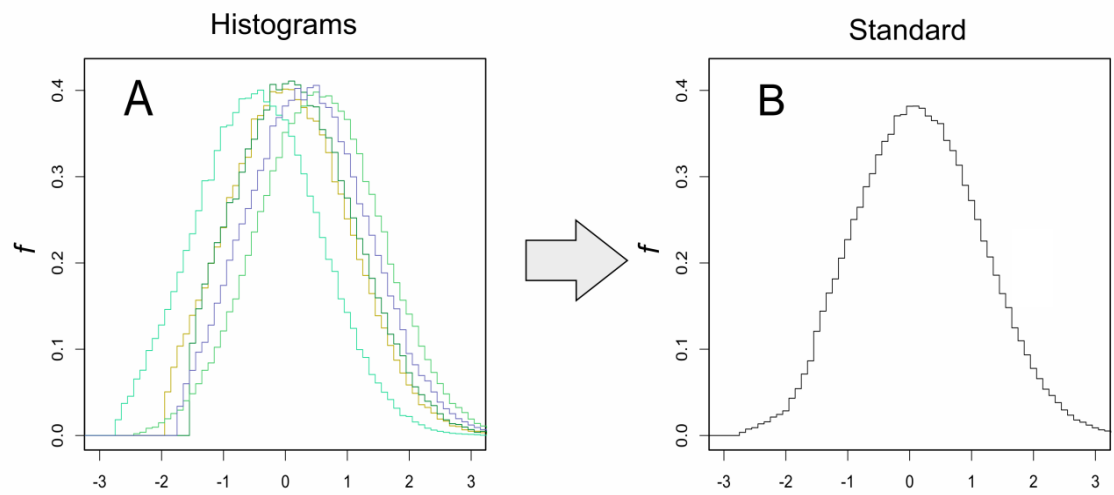

**Fig. 2 Equalization of data using RMA (model)**

**A.** Histograms of raw data may vary among measurements. **B.** A standard distribution could be estimated as averages of quantiles among measurements; the median of the top values, the median of the second values, etc. RMA replaces all the quantiles of the measurements with this standard, thus equalizing the data distributions.

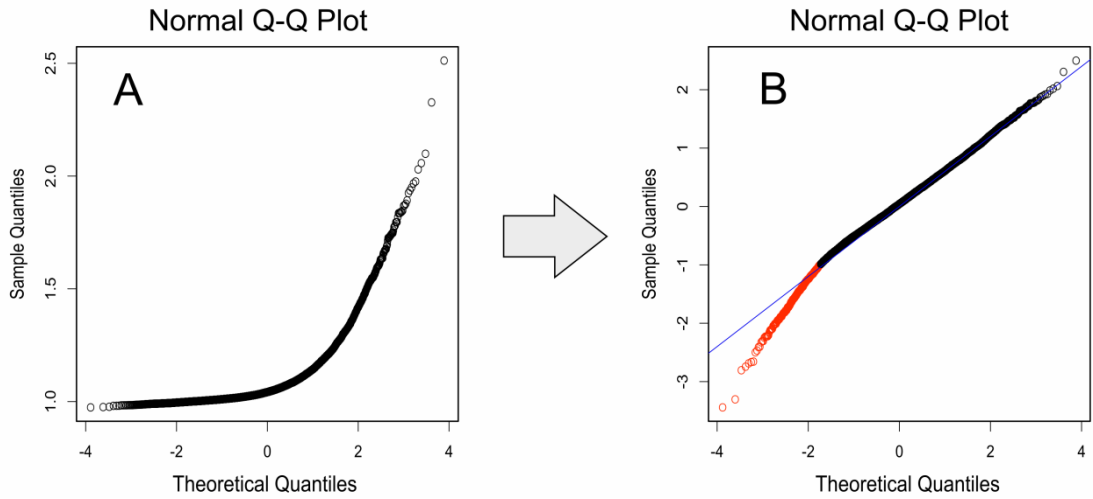

**Fig. 3 Parametric normalization (model)**

The distribution of data can be observed accurately using a quantile–quantile (Q–Q) plot. Logarithms of sorted raw data are compared against the theoretical values of the normal distribution, showing conflict with the model (**A**). However, if the transcriptome is distributed lognormally, a parameter for the hybridization background can be estimated by curve fitting (**B**). The lowest signal levels that are affected by noise do not coincide with the model (red). The two parameters of the normal distribution, position and scale, are found as the intercept and slope of the linear proportional relationship, respectively.

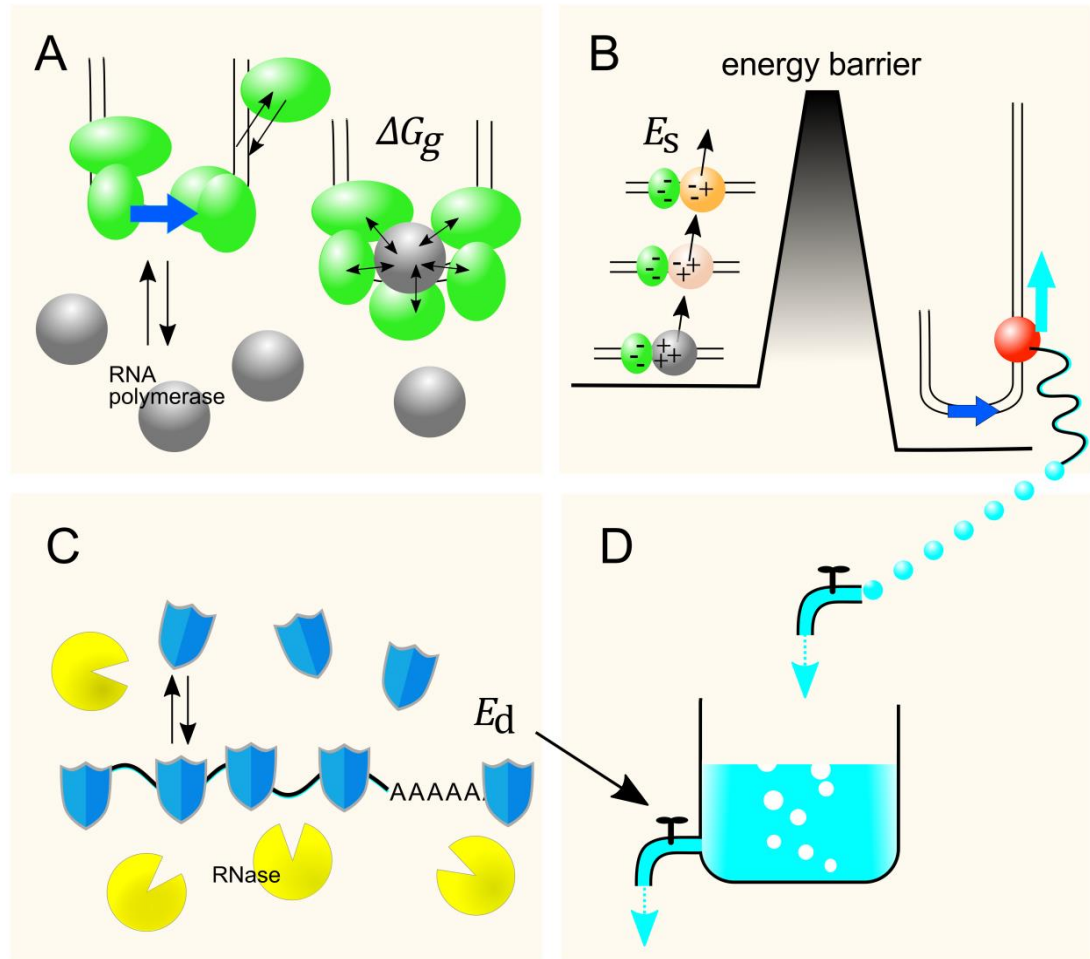

**Fig. 4 Thermodynamic model of transcriptome formation**

In the nucleus, active regulators occasionally attach to and detach from certain DNA motifs of the genome. These attachment and detachment actions are in equilibrium, and the stability of the binding is determined by the affinity of the regulator to the motif. Among the bound regulators, transcription factors may alter the stability of the binding of RNA polymerase to the promoter of a gene (**A**, green), while enhancer binding proteins may alter the frequency of activation of the polymerase (**B**). The manner in which the regulator acts is determined by the character of the regulator and the physical position at which the binding is occurring. The stability of RNA polymerase can be estimated by the sum total of the Gibbs free energy provided by the regulators (**A**, double-headed arrows). Moreover, the activation function can be expressed by the Arrhenius activation energy obtained by the sequential phosphorylation of the polymerase (**B**). Both functions are additive and proportional to the activity concentrations of the factors, as follows:

$$(\Delta G_g \text{ or } E_g) = c_{1 \text{ or } 2} \sum K_{g,p,i} k_{g,p,i} [\text{regulator}_i],$$

where $g$ specifies the gene, $p$ specifies the position of binding, and $i$ specifies the regulator protein. $K$ is the equilibrium constant of binding of the regulator and $k$ is a factor for the function of the regulator; it should be noted that both are recorded in the genome by the nucleotide sequence. The $c_{1-5}$ values are the constants that are common to a cell. According to the Arrhenius equation, the velocity of RNA synthesis from the gene can be determined using the following energy terms:

$$v_{s_g} = c_3 \exp\left(\frac{-\Delta G_g}{RT}\right) \times \exp\left(\frac{-E_{s_g}}{RT}\right), \tag{1}$$

where $R$ is the gas constant and $T$ is the absolute temperature.

In the cytoplasm, an RNA molecule is degraded spontaneously and has a specific half-life. RNA is attached to many proteins, which control its half-life (**C**). The Arrhenius activation energy for the degradation of gene $g$'s transcripts would be a function of the activity concentration of the RNA-binding protein $i$, as follows:

$$E_{d_g} = c_4 \sum K_{g,p,i} k_{g,p,i} [\text{RNAbinding}_i].$$

Using the energy, the velocity of degradation can be expressed as

$$v_{d_g} = -c_5 [\text{mRNA}_g] \exp\left(\frac{-E_{d_g}}{RT}\right). \tag{2}$$

The concentration of a transcript is determined by the pseudo-equilibrium of synthesis and degradation; this could be modeled as pouring water into a bucket with a hole (**D**). Hence,

$$v_{s_g} = -v_{d_g}. \tag{3}$$

From equations (1–3), the concentration is expressed, with the energies given by the factors and motifs as:

$$[\text{mRNA}_g] = c_3/c_5 \times \exp\left(\frac{-\Delta G_g - E_{s_g} + E_{d_g}}{RT}\right). \tag{4}$$

Eq. 4 shows that the molar concentration of a transcript in the cytosol is determined by the sum of the three categories of energies, each of which is the sum of the activities of various regulators. According to the central limit theorem, the sums of independent and identically distributed random numbers are normally distributed. The activities would satisfy this condition; hence, the concentrations will be lognormally distributed, as has been confirmed by each microarray and RNA-seq dataset.